\documentclass[conference]{IEEEtran}
\IEEEoverridecommandlockouts

\usepackage{cite}
\usepackage{amsmath,amssymb,amsfonts}
\usepackage{algorithmic}
\usepackage{graphicx}
\usepackage{textcomp}
\usepackage{xcolor}
\usepackage{orcidlink}
\usepackage{adjustbox}
\usepackage{tikz}
\usepackage{colortbl}
\usepackage{pgfplots}
\pgfplotsset{compat=newest}
\usepackage{pgfplotstable}

\definecolor{Gray}{gray}{0.85}
\definecolor{red_cool}{rgb}{0.5, 0.0, 0.0}

\usepackage[nolist, nohyperlinks]{acronym}
\def\BibTeX{{\rm B\kern-.05em{\sc i\kern-.025em b}\kern-.08em
    T\kern-.1667em\lower.7ex\hbox{E}\kern-.125emX}}

\begin{document}

\begin{acronym}
\acro{AE}{autoencoder}
\acro{ASV}{automatic speaker verification}
\acro{BCE}{binary cross-entropy}
\acro{CM}{countermeasure}
\acro{CNN}{convolutional neural network}
\acro{COTS}{commercial-off-the-shelf}
\acro{CQCC}{constant-Q cepstral coefficient}
\acro{CRNN}{convolutional recurrent neural network}
\acro{DNN}{deep neural network}
\acro{DF}{deepfake}
\acro{EER}{equal error rate}
\acro{ELU}{exponential linear unit}
\acro{FCN}{fully convolutional network}
\acro{FFT}{fast Fourier transform}
\acro{GMM}{Gaussian mixture model}
\acro{GRU}{gated recurrent unit}
\acro{ID}{identifier}
\acro{IoT}{Internet of things}
\acro{LA}{logical access}
\acro{LSTM}{long short-term memory}
\acro{MLP}{multilayer perceptron}
\acro{MVDR}{minimum variance distortionless response}
\acro{MWF}{multichannel Wiener filtering}
\acro{ReMASC}{Realistic Replay Attack Microphone Array Speech}
\acro{RNN}{recurrent neural network}
\acro{ROC}{receiving operating characteristic}
\acro{TTS}{text-to-speech}
\acro{PA}{Physical access}
\acro{PFA}{probability of false alarm}
\acro{SOTA}{state-of-the-art}
\acro{SRP-PHAT}{steered response power with phase transform}
\acro{STFT}{short-time Fourier transform}
\acro{VA}{voice assistant}
\acro{VC}{voice conversion}
\end{acronym}

\title{Multi-Channel Replay Speech Detection using Acoustic Maps
\thanks{}
}

\author{\IEEEauthorblockN{Michael~Neri~\orcidlink{0000-0002-6212-9139}, Tuomas~Virtanen~\orcidlink{0000-0002-4604-9729}}
\IEEEauthorblockA{\textit{Faculty of Information Technology and Commmunication Sciences}, \\ \textit{Tampere University}, Tampere, Finland}
\IEEEauthorblockA{\{michael.neri, tuomas.virtanen\}@tuni.fi}
}

\maketitle

\begin{abstract}
Replay attacks remain a critical vulnerability for automatic speaker verification systems, particularly in real-time voice assistant applications. In this work, we propose acoustic maps as a novel spatial feature representation for replay speech detection from multi-channel recordings. Derived from classical beamforming over discrete azimuth and elevation grids, acoustic maps encode directional energy distributions that reflect physical differences between human speech radiation and loudspeaker-based replay. A lightweight convolutional neural network is designed to operate on this representation, achieving competitive performance on the ReMASC dataset with approximately $6$k trainable parameters. Experimental results show that acoustic maps provide a compact and physically interpretable feature space for replay attack detection across different devices and acoustic environments.
\end{abstract}

\begin{IEEEkeywords}
Replay attack, Physical Access, Beamforming, Spatial Audio, Voice anti-spoofing, Acoustic maps.
\end{IEEEkeywords}

\section{Introduction}
Recently, \acp{VA} have become a central interface for human--machine interaction, exploiting speech as a biometric trait for user authentication and authorization~\cite{Huang_IoTJ_2022}. In practical deployments, \acp{VA} operate in real time to control \ac{IoT} devices and to transmit sensitive information, making timely and reliable spoofing detection a critical requirement. However, \ac{ASV} systems remain vulnerable to a wide range of audio-based attacks~\cite{Liu_TASLP_2023}.

Among these, \ac{LA} attacks manipulate speech content or speaker identity using \ac{TTS} or \ac{VC} techniques, while compression- or quantization-induced artifacts can further obscure such manipulations, leading to so-called \ac{DF} attacks~\cite{Liu_TASLP_2023}. \ac{PA} attacks, instead, aim to deceive the \ac{ASV} system at the microphone level~\cite{Kinnunen_ICASSP_2017}. In this setting, an adversary may either imitate the target speaker (impersonation attack)~\cite{Huang_IoTJ_2022} or replay a previously captured recording using a loudspeaker (replay attack)~\cite{Gong_SPL_2020}. In this work, we focus on replay attacks, as speech is inherently easy to capture in everyday environments and can be replayed with minimal effort~\cite{Delac_ISEM_2004}. Moreover, existing \ac{ASV} systems often struggle to reliably discriminate between genuine and replayed speech, even when using commodity off-the-shelf devices such as smartphones and loudspeakers~\cite{Neri_2025_ARXIV, Gong_Interspeech_2019}.

From a physical perspective, genuine speech and replayed speech are generated by fundamentally different sound production mechanisms: human speech originates from a complex vocal tract excitation, whereas replayed speech is emitted by an electro-acoustic transducer with its own frequency response, directivity, and radiation characteristics. These differences affect not only the spectral content of the signal but also its spatial propagation and interaction with the environment. In addition, the sound is emitted from a physical object, which can be either a human talker or a loudspeaker, so it does not originate from a single point but rather from a larger body with spatially distributed acoustic properties. Motivated by this observation, we investigate whether acoustic maps, capturing spatial and multi-channel sound field information, can be leveraged to distinguish between genuine and replayed speech excerpts in real time. By analyzing how sound is spatially produced and received across microphone arrays, we aim to assess whether these physical differences can be reliably exploited for replay attack detection.

\begin{figure*}[t!]
    \centering
    \centerline{\includegraphics[width=0.84\linewidth]{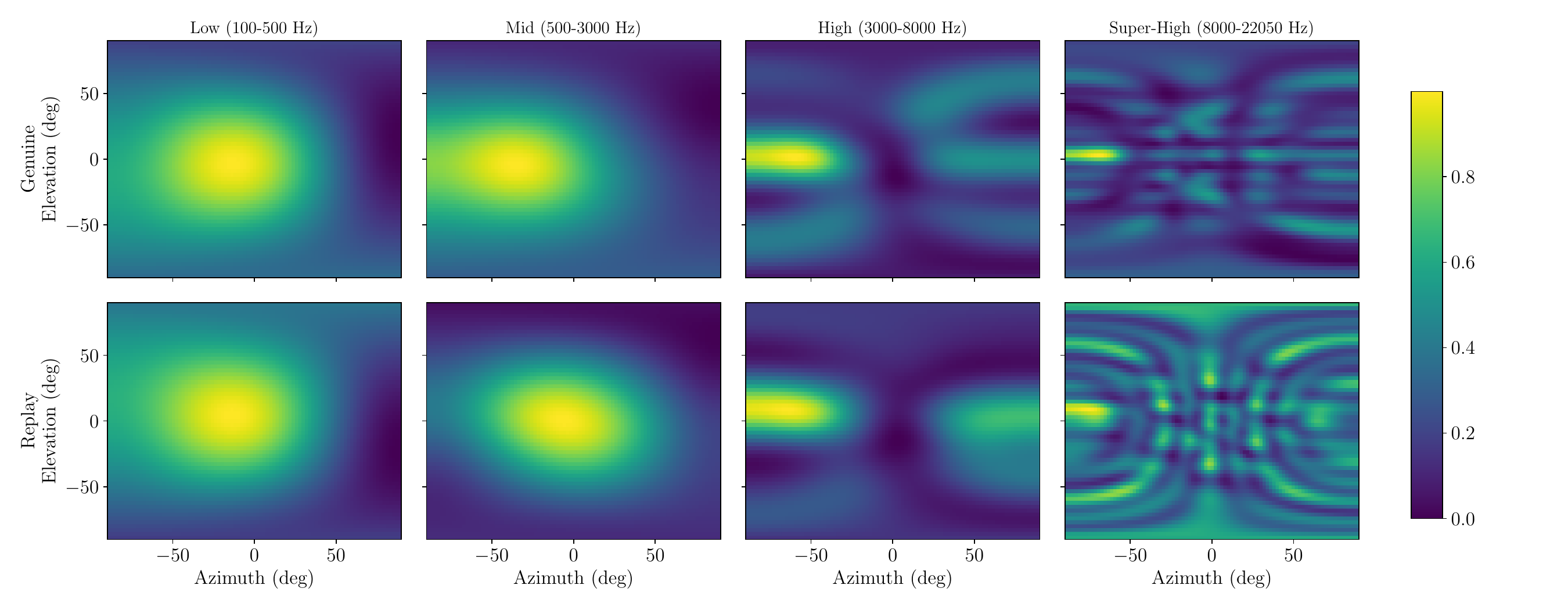}}
    \caption{Spatial distribution of acoustic maps from delay-and-sum beamformer across azimuth and elevation angles for two utterances from the ReMASC dataset, recorded using device $\mathrm{D}3$ with $6$ microphones arranged in a hexagonal shape. The top row corresponds to the genuine sample \textit{1264222.wav}, and the bottom row to the replay sample \textit{1380536001.wav} from the same indoor environment and positions. Each column represents a distinct frequency band: Low ($100–500$ Hz), Mid ($500–3000$ Hz), High ($3000–8000$ Hz), and Super-High ($8000–22050$ Hz). The color scale indicates normalized acoustic intensity.}
    \label{fig:AMGenuine_Replay}
\end{figure*}

The contributions of this work are as follows:
\begin{itemize}
    \item We introduce acoustic maps derived from classical beamforming as a spatial feature representation for replay speech detection, explicitly encoding directional energy distributions that reflect differences between human speech radiation and loudspeaker-based replay. 
    \item We design a compact convolutional neural network tailored to acoustic maps, achieving competitive replay detection performance on the \ac{ReMASC} dataset with approximately $6$k trainable parameters.
    \item We  evaluate the proposed approach under both environment-dependent and environment-independent conditions, analyzing robustness across different microphone arrays, beamformers, and unseen acoustic environments, and highlighting the strengths and limitations of spatial representations for replay detection.
\end{itemize}

The work is organized as follows: Sec.~\ref{sec:related} includes a literature review about replay speech detection, encompassing both traditional and learning-based approaches; Sec.~\ref{sec:method} describes the novel feature set and the \ac{CNN} used for detection; Sec~\ref{sec:results} illustrates the experimental results on real data and the comparison with prior works whereas Sec.~\ref{sec:conclusions} draws the conclusions.

\section{Related works}\label{sec:related}

In the literature, several works have tried to mitigate the replay speech attack by providing single-channel speech datasets, such as RedDots~\cite{Kinnunen_ICASSP_2017}, ASVSpoof2017 \ac{PA}~\cite{kinnunen2017asvspoof}, ASVSpoof2019~\cite{todisco2019asvspoof} \ac{PA}, and ASVSpoof2021 \ac{PA}~\cite{Liu_TASLP_2023}. Building upon these datasets, recent works in \ac{SOTA} focused on hand-crafted single-channel features with simple classifiers (such as \acp{CNN} and \acp{GMM}) after a beamforming phase (TECC~\cite{Kotta_2020_APSIPA}, CTECC~\cite{Acharya_2021_ICASSP}, and ETECC~\cite{Patil_2022_CSL}). However, all these methods suffer from generalization capabilities, i.e., changing the acoustic properties on the test set yields random guess predictions.  

However, for speech enhancement and separation tasks, microphone arrays are usually employed in \ac{ASV} systems to exploit spatial information and improve audio quality~\cite{omologo2001speech}. Moreover, multi-channel data can be beneficial for detecting the replay detection for several reasons: (i) multi-channel recordings encompass audio spatial cues that can help the detection~\cite{Gong_Interspeech_2019, Gong_SPL_2020, Neri_2025_EUSIPCO}, and (ii) such spatial information cannot be easily counterfeited by an attacker, differently to single-channel data where temporal and frequency cues can be manipulated to fool an \ac{ASV} system~\cite{zhang2016voicelive}. 

Despite these advantages, progress in multi-channel replay detection has been limited by the lack of suitable datasets and models specifically designed to capture and exploit spatial information. To date, \ac{ReMASC}~\cite{Gong_Interspeech_2019} remains the only publicly available dataset providing real multi-channel recordings for replay attack detection, encompassing different microphone arrays, playback devices, and acoustic environments. The baseline of this dataset was CQCC-GMM~\cite{Gong_Interspeech_2019}, which is a \ac{GMM} classifier trained on \acp{CQCC}. They also proposed a \ac{CRNN} which acts as both a learnable and adaptive time-domain beamformer with a \ac{CRNN} replay attack classifier~\cite{Gong_SPL_2020}. In~\cite{Neri_2025_ARXIV} a \ac{CRNN} replay detector with a learnable time-frequency beamformed was proposed, achieving SOTA performance on ReMASC. To partially alleviate this data scarcity,  an acoustic simulation framework was recently proposed in~\cite{neri2025acoustic} to generate synthetic genuine and replay multi-channel recordings.

\section{Acoustic Map-based Replay Detector}\label{sec:method}
In this section we describe how we compute the acoustic maps from a multi-channel recording using the delay-and-sum beamformer. Then, we describe the lightweight \ac{CNN}, which is responsible for distinguishing between genuine and replay speeches, that is, if the sound is emitted from a real talker or a loudspeaker.

\subsection{Beamforming-based Acoustic Maps}
Let $\mathbf{x}\in\mathbb{R}^{N\times T_s}$ denote a multi-channel audio segment with $T_s$ samples recorded from $N$ microphones.  
Each channel is first pre-processed using the \ac{STFT} with window length $N_{\mathrm{FFT}}$ and hop size $H$:
\begin{equation}
\mathbf{X}_i(f,t)=\mathrm{STFT}\{\mathbf{x}_i\}, 
\qquad 
\mathbf{X}\in\mathbb{C}^{N\times F\times T},
\end{equation}
where $F$ is the number of frequency bins and $T$ is the number of frames. The microphone geometry is encoded by
\begin{equation}
\mathbf{P}=
\begin{bmatrix}\mathbf{p}_1^\top\\ \vdots\\ \mathbf{p}_N^\top\end{bmatrix}
\in\mathbb{R}^{N\times 3},
\end{equation}
where $\mathbf{p}_i$ is the vector defining the position of the $i$-th microphone in the space.

For computing the acoustic maps, we define a discrete spatial grid sampled from a set of azimuth ($\mathrm{Az}$) and elevation ($\mathrm{El}$) angles as
\begin{equation}
\mathcal{G}=\{(\alpha_k,\beta_l)\mid \alpha_k\in\mathrm{Az},~\beta_l\in\mathrm{El}\},
\end{equation} 
where $\alpha_k$ and $\beta_l$ denote a single point of the grid with a fixed azimuth and elevation angles, respectively. Specifically, the azimuth and elevation angle sets are defined as uniformly sampled angular grids. The azimuth set $\mathrm{Az}$ consists of $91$ discrete angles uniformly 
spanning the interval $[-90^\circ,\,90^\circ]$, while the elevation set $\mathrm{El}$ consists of $41$ discrete angles uniformly spanning the same interval.

For each grid point $(\alpha,\beta)$, the corresponding unit direction vector is
\begin{equation}
\mathbf{u}(\alpha,\beta)=
\begin{bmatrix}
\cos\beta\cos\alpha\\
\cos\beta\sin\alpha\\
\sin\beta
\end{bmatrix},
\end{equation}
and the plane-wave steering vector is 
\begin{equation}
\mathbf{a}(\omega,\alpha,\beta)
=\exp\!\left(-j\frac{\omega}{c}\,\mathbf{P}\,\mathbf{u}(\alpha,\beta)\right)
\in\mathbb{C}^{N},
\end{equation}
where $c = 343\, \mathrm{m/s}$ is the speed of sound, $\omega = 2\pi f_{\mathrm{Hz}}$ denotes the angular frequency (rad/s) 
corresponding to a physical frequency $f_{\mathrm{Hz}}$, and $j = \sqrt{-1}$. Using this steering vector, a narrowband pseudo-power response is computed for each $(f,t)$ as
\begin{equation}
M(f,t,\alpha,\beta)
= \left|\mathbf{a}(f,\alpha,\beta)^{\mathrm{H}}\,\mathbf{X}(f,t)\right|^{2},
\end{equation}
where $\mathbf{X}(f,t)$ is a vector consisting of all the pre-processed channels. To obtain a more compact representation, the maps are time averaged and the frequency bins are grouped into $K$ disjoint bands, where each $B_m$ is a set consisting of frequency indices of the $m$-th band
\begin{equation}
B_m\subseteq\{1,\ldots,F\},\qquad m=1,\ldots,K,
\end{equation}
and the band- and time-averaged acoustic map is computed as
\begin{equation}
M_m(a,e)
=\frac{1}{|B_m|\,T_s}
\sum_{f\in B_m}
\sum_{t=1}^{T_s}
M(f,t,\alpha_a,\beta_e),
\end{equation}
where $a\in\{1,\ldots,A\}$ and $e\in\{1,\ldots,E\}$ index the azimuth and elevation grid points, so that $\alpha_a$ and $\beta_e$ are the corresponding angles. The final representation is a 3-D acoustic-map tensor $\mathbf{M}\in\mathbb{R}^{K\times A\times E}$ with $A = |\mathrm{Az}| = 91$ as the number of azimuths and $E = |\mathrm{El}| = 41$ as the number of elevations, encoding sound radiation patterns.  Moreover, the acoustic map can encompass directional energy distributions that capture array-manifold structure, direct-path cues, and early reflections, all of which are informative for discriminating genuine and replayed signals. An example of the computed acoustic maps using delay-and-sum on genuine and replay recordings are depicted in Fig.~\ref{fig:AMGenuine_Replay}.

Although this subsection explained the computation using the delay-and-sum beamforming, acoustic maps are not tied to a specific spatial processing method and can also be computed using alternative approaches, such as \ac{MVDR} beamforming or \ac{SRP-PHAT}, which are evaluated in later experiments.

\subsection{Convolutional Neural Network Classifier}
The computed acoustic maps are then fed to a neural network that is responsible for classifying between genuine and replay recordings. Specifically, the proposed lightweight \ac{CNN} operates on acoustic maps $\mathbf{M} \in \mathbb{R}^{K \times A \times E}$. The architecture is designed to balance representational capacity and computational efficiency, making it suitable for data-limited and resource-constrained scenarios. The network is built around a repeated convolutional block composed of a depthwise separable 2D convolution operating over $A$ and $E$ dimensions, followed by batch normalization, an \ac{ELU} activation function~\cite{Djork_2016_ICLR}, and $(2 \times 2)$ max pooling. Depthwise separable convolutions substantially reduce the number of trainable parameters and computational cost compared to standard convolutions~\cite{Neri_LSENS_2024}, while still capturing local spectro-temporal patterns in the acoustic maps. This block is applied three times with progressively increasing channel dimensionality and decreasing kernel size, namely $(K \rightarrow 8, k=5)$, $(8 \rightarrow 16, k=3)$, and $(16 \rightarrow 32, k=3)$, enabling hierarchical feature extraction with controlled model capacity. After these three blocks, a final depthwise separable 2D convolution with $32$ input and output channels and kernel size $3$ is applied, followed by batch normalization and ELU~\cite{Djork_2016_ICLR} activation, without further pooling. Early spatial downsampling in the repeated blocks reduces the resolution of intermediate representations, limiting sensitivity to noise and fine-grained variations and improving generalization. The resulting feature maps are projected to $2$ channels and flattened. The flattened representation is processed by a shallow fully connected classification head composed of a \ac{MLP} with $110$ neurons and $32$ hidden units, followed by batch normalization and ELU~\cite{Djork_2016_ICLR} activation, and a final linear layer mapping to a $2$-dimensional output for the classification between genuine and replay classes. The projection to a low-dimensional embedding and the shallow classifier further constrain model capacity and mitigate overfitting.

Overall, the architecture contains approximately $6000$ trainable parameters and provides a compact yet effective alternative to deeper or attention-based models for acoustic scene analysis. The \ac{DNN} has been trained with the categorical cross-entropy loss with sofmax nonlinearity. In addiction, we employed MixUp augmentation~\cite{zhang2018mixup} with $\alpha = 0.05$. 

\section{Experiments}\label{sec:results}
\subsection{Dataset}
\ac{ReMASC}~\cite{Gong_Interspeech_2019} is selected as it is the only publicly available dataset providing synchronized multi-channel recordings acquired with devices featuring different numbers of microphones and array geometries for this task. The dataset comprises recordings from four \ac{ASV} microphone arrays, denoted as ${\mathrm{D}1, \mathrm{D}2, \mathrm{D}3, \mathrm{D}4}$, equipped with $2$, $4$, $6$, and $7$ omnidirectional microphones, respectively. $\mathrm{D}1$ and $\mathrm{D}2$ are linear microphone arrays, while $\mathrm{D}3$ and $\mathrm{D}4$ are arranged in a hexagonal configuration, with $\mathrm{D}4$ additionally including a central microphone. In total, the dataset includes $9{,}240$ genuine and $45{,}472$ replay audio samples. Devices $\mathrm{D}1$, $\mathrm{D}2$, and $\mathrm{D}3$ operate at a sampling rate of $44.1$~kHz, while $\mathrm{D}4$ records at $16$~kHz. The bit depth is $16$ bits for all devices except $\mathrm{D}3$, which uses $32$ bits. \ac{ReMASC} features speech from $55$ speakers with diverse gender and vocal characteristics, recorded across four different acoustic environments: an outdoor scenario (Env-A), two enclosed spaces (Env-B and Env-C), and a moving vehicle (Env-D). Genuine speech is captured using two spoofing microphones and subsequently replayed through four playback devices of varying quality, enabling the simulation of realistic replay attack scenarios under diverse acoustic conditions.

To assess the performance of our approach and directly compare with \ac{SOTA} models, \ac{EER} metric is used following the same train-test split provided by the authors of the dataset~\cite{Gong_Interspeech_2019}. A separate model is trained for each microphone array in the dataset. We perform five independent runs to compute the $95\%$ mean confidence intervals. All four environments are present in the training and testing sets. However, we perform generalization capabilities to unseen environments later in the experiments, i.e., when one of the them is removed from the training set.  

\subsection{Results}
The results in Table~\ref{tab:resultsComplMicWise} show that the proposed acoustic-map-based approach achieves diverse performance across microphone arrays. Detection accuracy is clearly influenced by the number of microphones and array geometry: arrays with more microphones ($\mathrm{D}3$ and $\mathrm{D}4$) consistently yield lower \acp{EER} compared to $\mathrm{D}1$ and $\mathrm{D}2$. This trend is expected, as larger arrays provide richer spatial sampling of the sound field, improving the reliability of directional energy cues encoded in the acoustic maps. Compared to learning-based multichannel baselines, the proposed method does not uniformly outperform state-of-the-art approaches; however, from a model-complexity perspective, the proposed approach is highly efficient in terms of number of learnable parameters: with approximately $6$k parameters, it is significantly lighter than M-ALRAD ($\approx 300$k) and the \ac{ReMASC} \ac{CRNN} baseline ($ \approx 1$M). This supports the claim that acoustic maps, combined with a shallow CNN, offer a favorable trade-off between performance and computational cost. The results indicate that acoustic maps are informative for replay detection, but their effectiveness is constrained when spatial resolution is limited, as in small arrays.

\begin{table}[ht!]
\caption{Microphone-wise comparison of \ac{EER} ($\%$) with \ac{ReMASC}  baseline.}
\label{tab:resultsComplMicWise}
\centering
\adjustbox{max width=0.5\textwidth}{%
    \begin{tabular}{c|cccc}
    \hline \hline 
    Methods & $\mathrm{D}1$ & $\mathrm{D}2$ & $\mathrm{D}3$ & $\mathrm{D}4$  \\ 
    \hline
    NN-Single~\cite{Gong_SPL_2020} & $16.6$ & $23.7$ & $23.7$ & $27.5$ \\
    NN-Dummy Multichannel~\cite{Gong_SPL_2020} & $16.0$ & $23.2$ & $24.5$ & $25.2$ \\
    NN-Multichannel~\cite{Gong_SPL_2020} & $14.9$ & $15.4$ & $16.5$ & $19.8$ \\
    ALRAD~\cite{Neri_2025_ARXIV} & $5.5 \pm {\scriptstyle 2.6}$ & $11.9 \pm {\scriptstyle 2.2}$ & $19.5 \pm {\scriptstyle 2.2}$ & $21.7 \pm {\scriptstyle 2.1}$ \\
    M-ALRAD~\cite{Neri_2025_ARXIV} & $\mathbf{5.2} \pm {\scriptstyle 1.2}$ & $\mathbf{10.0} \pm {\scriptstyle 0.9}$ & $10.4 \pm {\scriptstyle 2.1}$ & $\mathbf{14.2} \pm {\scriptstyle 0.8}$ \\
    \hline
    \rowcolor{Gray} Acoustic maps & $21.6 \pm {\scriptstyle 1.0}$ & $19.9 \pm {\scriptstyle 2.8}$ & $\mathbf{10.1} \pm {\scriptstyle 2.8}$ & $19.7 \pm {\scriptstyle 2.8}$  \\
    \hline \hline
    \end{tabular}
    }
\end{table}

\subsection{Analysis on the choice of beamformer}

As previously mentioned, different spatial processing methods can be employed to compute the acoustic maps. We studied the performance of delay-and-sum beamforming, MVDR beamforming, and SRP-PHAT in Table~\ref{tab:beamformers}, which highlights the impact of the chosen method to generate acoustic maps on replay detection accuracy. 

Overall, the results indicate that the choice of spatial processing technique influences performance in a device-dependent manner. Delay-and-sum beamforming achieves competitive performance across all microphone arrays and provides a stable baseline, particularly for arrays with a lower number of microphones, e.g. $\mathrm{D}1$. \ac{MVDR} beamforming yields comparable results and, in some cases, slightly improved performance (e.g., for $\mathrm{D}3$ and $\mathrm{D}4$), but may be more sensitive to estimation errors in the spatial covariance matrices, especially for short or noisy utterances. \ac{SRP-PHAT} generally exhibits higher \acp{EER}, which may be attributed to its sensitivity to reverberation and noise, potentially leading to less reliable spatial energy patterns for this task.

These results suggest that while more advanced spatial processing methods can be beneficial, simpler approaches such as delay-and-sum remain effective and robust for generating acoustic maps in replay detection scenarios.

\begin{table}[ht!]
\caption{\ac{EER} ($\%$) changing the type of beamformer for computing the acoustic maps.}
\label{tab:beamformers}
\centering
\adjustbox{max width=0.5\textwidth}{%
    \begin{tabular}{c|cccc}
    \hline \hline 
    Methods & $\mathrm{D}1$ & $\mathrm{D}2$ & $\mathrm{D}3$ & $\mathrm{D}4$  \\ 
    \hline
    Delay-and-sum & $\mathbf{21.6} \pm {\scriptstyle 1.0}$ & $19.9 \pm {\scriptstyle 2.8}$ & $10.1 \pm {\scriptstyle 2.8}$ & $19.7 \pm {\scriptstyle 2.8}$ \\
    MVDR & $32.4 \pm {\scriptstyle 1.3}$ & $20.6 \pm {\scriptstyle 1.5}$ & $\mathbf{9.9} \pm {\scriptstyle 2.7}$ & $\mathbf{17.6} \pm {\scriptstyle 1.6}$ \\
    SRP-PHAT & $31.4 \pm {\scriptstyle 1.6}$ & $\mathbf{19.4} \pm {\scriptstyle 2.5}$ & $12.5 \pm {\scriptstyle 5.6}$ & $21.2 \pm {\scriptstyle 2.2}$ \\
    \hline \hline
    \end{tabular}
    }
\end{table}

\begin{figure}[ht!]
    \centering
    \centerline{\includegraphics[width=\linewidth]{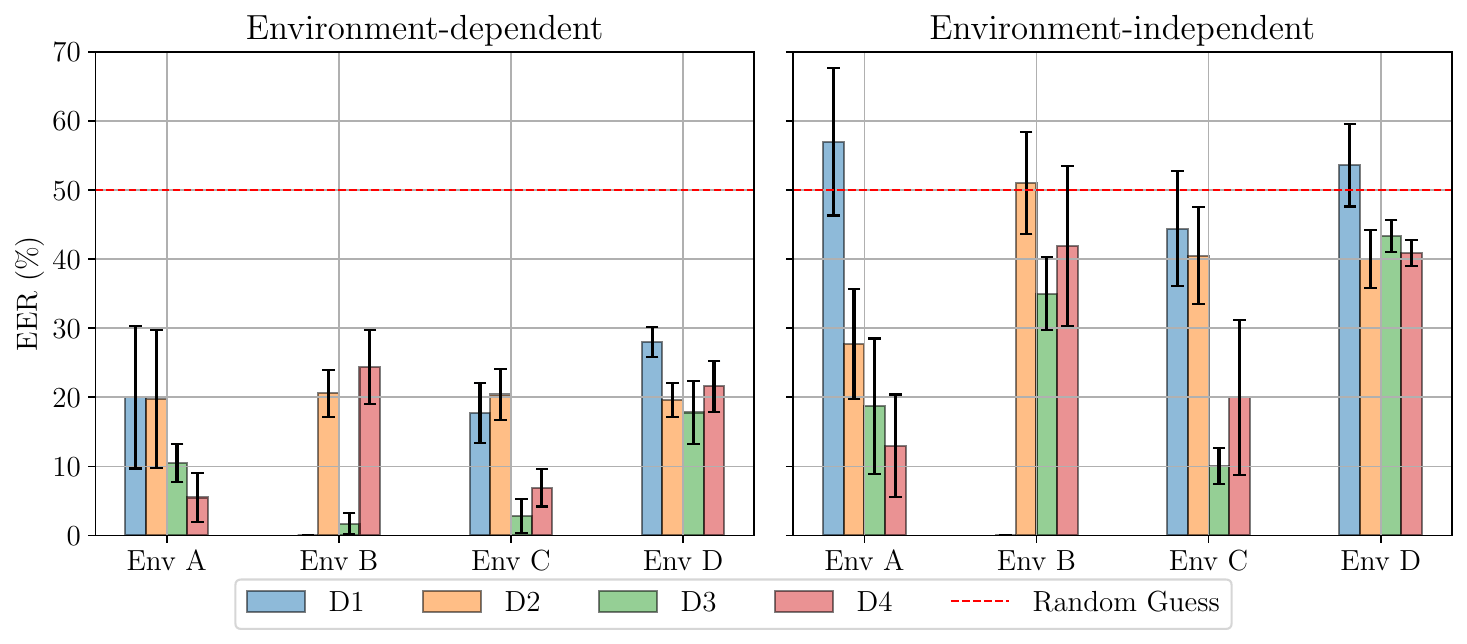}}
    \caption{Microphone-wise performance in both generalization scenarios.}
    \label{fig:eer}
\end{figure}

\subsection{Analysis on generalization}
As done in previous works~\cite{Neri_2025_ARXIV, Acharya_2021_ICASSP}, we provide the performance of acoustic maps on ReMASC in a \textit{environment-dependent} and \textit{-independent} scenarios in Table~\ref{tab:resultsEnv}. Specifically, in the \textit{environment-dependent} setting, training data encompasses all the environments, and the model is tested on the same environments. Differently, the \textit{environment-independent} setting is designed to assess robustness to unseen conditions: the model is trained using data from three environments and evaluated on the remaining fourth unseen environment. This procedure is repeated for each environment.

In the \textit{environment-dependent setting}, acoustic maps achieve reasonable performance, although they do not consistently outperform handcrafted-feature-based methods such as TECC~\cite{Kotta_2020_APSIPA} or M-ALRAD~\cite{Neri_2025_ARXIV}. This suggests that while spatial information is useful, it does not fully replace carefully designed spectral features under matched conditions.

In the more challenging \textit{environment-independent} setting, performance degradation is observed for all methods, including the proposed approach. Acoustic maps exhibit limited robustness to unseen environments, with \acp{EER} increasing substantially, particularly for arrays with fewer microphones. This highlights a key limitation of the current representation: spatial patterns captured by acoustic maps remain sensitive to environmental changes such as room geometry and reverberation. In addition, fixed frequency bands and static beamforming limit adaptability, particularly in unseen environments. This suggests that future improvements should focus on learning frequency-dependent or adaptive spatial representations to better capture replay-specific directivity cues. 

Nevertheless, the results confirm that acoustic maps generalize at a level comparable to other multichannel approaches, despite using a fixed, non-adaptive spatial representation and a lightweight classifier.

\begin{table}[ht!]
\caption{\acp{EER} ($\%$) for environment-dependent vs. environment-independent scenarios on ReMASC dataset.}
\label{tab:resultsEnv}
\centering
\adjustbox{max width=0.5\textwidth}{%
    \begin{tabular}{c|cccc}
    \hline \hline 
    Methods & Env-A & Env-B$^{*}$ & Env-C & Env-D  \\ 
    \hline
    \multicolumn{5}{c}{Environment-dependent} \\
    CQCC-GMM~\cite{Gong_Interspeech_2019} & $13.5$  & $17.4$  & $21.3$  & $22.1$  \\
    ETECC~\cite{Patil_2022_CSL} & $15.1$  & $33.8$  & $14.1$  & $10.4$ \\
    CTECC~\cite{Acharya_2021_ICASSP} & $13.0$  & $26.8$  & $9.9$  & $10.1$ \\
    TECC~\cite{Kotta_2020_APSIPA} & $13.4$ & $27.9$ & $10.3$ & $\mathbf{9.1}$ \\
    M-ALRAD~\cite{Neri_2025_ARXIV} & $\mathbf{8.1} \pm {\scriptstyle 2.3}$ &  $\mathbf{5.8} \pm {\scriptstyle 2.4}$  &  $\mathbf{7.5} \pm {\scriptstyle 2.6}$ & $14.0 \pm {\scriptstyle 2.0}$  \\
    \rowcolor{Gray} Acoustic maps & $13.9 \pm {\scriptstyle 4.0}$ &  $19.2 \pm {\scriptstyle 9.5}$  &  $11.9 \pm {\scriptstyle 3.7}$ & $21.7 \pm {\scriptstyle 2.2}$\\
    \hline
    \multicolumn{5}{c}{Environment-independent} \\
    CQCC-GMM~\cite{Gong_Interspeech_2019} & $19.9$  & $39.9$  & $34.6$  & $48.9$  \\
    ETECC~\cite{Patil_2022_CSL} & $29.0$  & $32.5$  & $30.0$  & $49.9$ \\
    CTECC~\cite{Acharya_2021_ICASSP} & $28.5$  & $34.7$  & $32.5$  & $50.0$ \\
    TECC~\cite{Kotta_2020_APSIPA} & $26.8$ & $35.4$ & $31.8$ & $50.0$ \\
    M-ALRAD~\cite{Neri_2025_ARXIV} & $\mathbf{13.8} \pm {\scriptstyle 2.8}$ & $\mathbf{20.3} \pm {\scriptstyle 3.1}$  & $\mathbf{15.1} \pm {\scriptstyle 5.5}$ & $\mathbf{24.2} \pm {\scriptstyle 2.4}$  \\
    \rowcolor{Gray} Acoustic maps & $27.1 \pm {\scriptstyle 7.1}$ &  $41.6 \pm {\scriptstyle 4.3}$  &  $28.3 \pm {\scriptstyle 7.4}$ & $43.2 \pm {\scriptstyle 2.0}$ \\
    \hline \hline
    \multicolumn{5}{l}{\scriptsize Env-B$^{*}$ does not encompass D1 genuine utterances due to hardware fault during data collection.} \\
    \end{tabular}
    }
\end{table}

\section{Conclusion}\label{sec:conclusions}
In this work, we proposed acoustic maps as a spatial feature representation for replay speech detection from multi-channel recordings. The proposed approach exploits classical spatial processing to project multi-channel audio onto discrete azimuth and elevation grids, encoding directional energy distributions that reflect physical differences between genuine human speech radiation and loudspeaker-based replay. Experimental results on the \ac{ReMASC} dataset demonstrate that acoustic maps provide a compact and physically interpretable representation for replay detection, achieving competitive performance with a lightweight convolutional neural network containing approximately $6000$ trainable parameters. At the same time, the results highlight limitations of the current formulation, particularly in environment-independent scenarios, where fixed frequency bands and static spatial processing reduce adaptability to unseen acoustic conditions.
Future work will focus on the implementation of a learning-based frequency band selector to better capture directivity cues from multi-channel mixtures.

\bibliographystyle{IEEEtran}
\bibliography{biblio}
\end{document}